# Stress chain analysis of soil particles in slope via persistent homology

Shengdong ZHANG[1]   Shihui YOU*[1]   Longfei CHEN[2]   Xiaofei LIU[2]

1. College of Mechanical and Electrical Engineering, Zaozhuang University 277160;

2. College of Civil Engineering and Mechanics, Xiangtan University 411105

**Abstract:** The relationship between the macroscopic response of the slope and the macrostructure of the force chain network under the action of the metal plate was studied by the particle discrete element method and the persistent homology. The particle accumulation model was used to simulate the instability process of slope under the continuous downward action of metal plate by the particle discrete element method. The macroscopic responses such as the total velocity vector of the two-dimensional slope deposit, the angle of the slip cracking surface when the slope is unstable, and the average velocity in the $y$-direction of the slope were studied. Then, the normal force chain undirected network model of the natural accumulation of slope stacking particles was constructed. Finally, the topological characteristics of the particle contact force chain network of the slope top were analyzed by the persistent homology method to obtain the barcode. Finally, the relationship between the instability evolution and the characteristics of persistent homology is established. This research provides a new method for the study of slope instability topology identification. Thus, the instability destruction of slope can be predicted effectively.

## 1 Introduction

Slope instability accidents widely exist in geotechnical engineering, environmental science, transportation and many other fields. Slope will lose stability and cause landslide due to rain erosion, earthquake and impact, which will cause huge economic loss to the society and cause huge life threat to the people. Therefore, the study of slope instability has very important theoretical and scientific significance. The main research methods of slope failure include limit equilibrium method, model experiment method and numerical analysis method. Numerical analysis method includes discrete element method [1], finite element method, discontinuous deformation analysis method and so on. The advantage of numerical analysis is that it has low cost, good repeatability, and easy implementation. The particle discrete element method (PDEM) does not need to set the constitutive relation of materials and the position of sliding surface, but directly defines the contact relationship between particles from the meso level. The calculation process is the adjustment process of the internal material of the slope to obtain the stable state. It does not require the deformation coordination and continuous displacement, and simplifies the calculation process. At present, scholars have done a lot of research on the stability of rock mass and slope by using discrete element method. Fakhimi, potyondy, backstron, etc. studied the deformation and failure characteristics, macro properties and micro characteristics of rock mass through simulated uniaxial compression test [2-5]. Based on PFC2D and EDEM software, many scholars such as Yang Bing, Zhang Xiaoxue and Yang Ling studied and discussed the macro response of slope failure process [6-10]. In recent years, persistent homology has made great progress in image processing and target recognition. Zhang Jingliang, Ju xianmeng [11] used the calculation method of persistent homology and simple complex homology group to classify and recognize images. By calculating the homology of the simple complex, the barcode is obtained, and the topological features and corresponding geometric structure information of the image are obtained based on the barcode. In this thesis, the particle discrete element method is used to study the stability of soil slope. A static load is applied to the upper part of the slope

*Corresponding author

**E-mail address:** 101434@uzz.edu.cn

by the downward pressure of metal plate to simulate the failure and instability process of the slope when there is load on the upper part of the slope, and the mechanical behavior of the slope top in different time periods is analyzed. Finally the topological analysis of the force chain network on the top of soil slope is carried out by the persistent homology theory.

## 2 Numerical Simulation Method and Network Analysis

### 2.1 Particle Discrete Element Method

Hertz Mindlin with bonding bonding model is used, in which particles are bonded together by a certain size of "viscous bond". When the maximum normal and tangential shear stress is reached, the viscous bond will break. The failure of the viscous bond leads to the failure of the slope. The standard Hertz-Mindlin contact model is used to solve the contact problem before the formation of particle bonding. The ratio of normal force to moment ($F_{n,t}/T_{n,t}$) of particles returns to 0 after the particles are bonded at time $t_{BOND}$, and the values of normal force and moment are updated at each calculation time step by the following equation:

$$\delta F_n = -v_n S_n A \delta t \quad (1)$$

$$\delta F_t = -v_t S_t A \delta t \quad (2)$$

$$\delta M_n = -\omega_n S_t J \delta t \quad (3)$$

$$\delta M_t = -\omega_t S_n \frac{J}{2} \delta t \quad (4)$$

$$A = \pi R_B^2 \quad (5)$$

$$J = \frac{1}{2} \pi R_B^4 \quad (6)$$

Where $A$ is the contact area; $S_n$ and $S_t$ are the normal stiffness and tangential stiffness respectively; $R_B$ is the radius of the "viscous bond"; $\delta_t$ is the time step; $v_n$ and $v_t$ are the normal and tangential velocities of particles; $\omega_n$ and $\omega_t$ are the normal and tangential angular velocities respectively. When the normal and tangential stresses exceed the preset threshold, the bond will break. Therefore, the maximum values of normal and tangential stresses are defined as follows:

$$\sigma_{\max} < \frac{-F_t}{A} + \frac{2M_t}{J} R_B \quad (7)$$

$$\tau_{\max} < \frac{-F_t}{A} + \frac{M_t}{J} R_B \quad (8)$$

### 2.2 Basic principles of persistent homology

Persistent homology is the introduction of persistence on the basis of homology, which is the generalization of homology. Homology is a static concept, while persistent homology is a changing and dynamic process. Homology calculation is carried out on simple complex, while persistent homology is calculated on filtered nested complex. Persistent homology contains many basic concepts, such as Betti number, simplex, simple complex and so on. Simple complex is the basis of persistent homology. With simple complex, we can construct homology group and calculate Betti number. By mapping the simple complex, a filter complex is obtained, and the corresponding bar code diagram is obtained. In the bar code diagram, those with short duration are noise, while those with longer duration are the real topological features. The bar code graph can directly reflect the Betti number of the complex, so we can get the real topological invariants of the data. Next, we introduce the concept of homology group.

**Homology Group**

Group is a kind of mathematical structure, which is used to describe all things with symmetry concept. And now it has developed into a complete mathematical system group theory. Group is a very common concept, it can be applied to many objects. If a high-dimensional simplex complex is regarded as a combination of all the lower dimensional simplex in accordance with some increasing law, the increasing sequence is called filtered nested complex. It is known that simplexes in complex *K* are connected with each other by faces, so nested complex is a sequence generated by nesting sub complex with each other

$$\phi \subset K^1 \subset K^2 \subset \cdots \subset K^{n-1} \subset K$$

Under the mapping of inclusion relations, a homology has a change process and a life cycle. Life cycle can be described in terms of Betti interval.

Homology Group: A nested complex *K*, The boundary operator corresponding to its *i*-th complex $K^i$ is $\partial_p^i$, group $C_p^i, Z_p^i, B_p^i$ and quotient group $H_p^i, \forall i$, $k>0$, so, the p-th homology group of $K^i$ is

$$H_p^{i,j} = Z_p^i / (B_p^j \cap Z_p^i)$$

$$b_p^{i,j} = \text{rank}\, H_p^{i,j}$$

The p-dimension Betti number of simplex $K^i$ is $b_p^{i,j} = \text{rank}\, H_p^{i,j}$

Betti number is an important invariant. Intuitively, $b_0$ is the number of connected parts, $b_1$ is the number of one-dimensional holes, $b_2$ is the three-dimensional holes surrounded by faces, and so on. For example, the Betti number of a torus should be $b_0 = 1$, $b_1 = 2$, $b_2 = 1$. It represents a connecting component, two one-dimensional holes and one three-dimensional hole. For persistent homology, betti interval can describe the change of homology group with $\varepsilon$. A *k*-dimensional Betty interval describes the range of $\varepsilon$ from the appearance to the disappearance of a *k*-dimensional hole. If we draw the betti interval in a two-dimensional coordinate system, we can get a visual description of continuous coherence.

## 3 Numerical simulation
### 3.1 Computational model
The discrete element software is used to simulate the process of slope failure. The discrete element model is shown in Figure 1. The slope model is composed of 1637 uniform particles with a diameter of 3mm. Fixed end constraints are applied to the left, right and bottom surface of the slope. At different positions of the slope top, A, B, C and D ($x_A$=-52.5mm, $x_B$=-62.5mm, $x_C$=-72.5mm, $x_D$=-82.5mm) at a certain speed along the Y axis Load the soil slope. The basic characteristic parameters and bond model parameters of soil granular materials are determined by compression test [12], as listed in Table 1 and table 2.

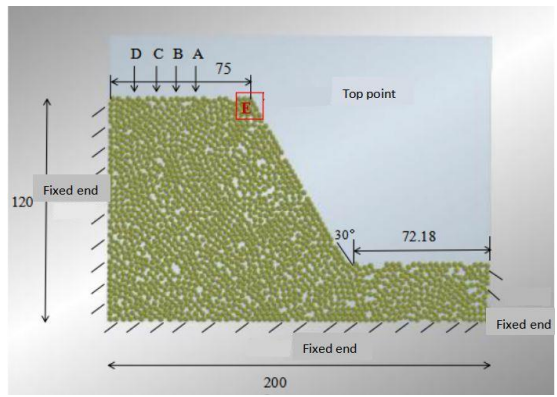

**Fig.1 Soil slope model**

**Tab.1 Granular material parameters**

| Destiny(kg m$^{-3}$) | 2000 | Poisson ratio | 0.35 |
|---|---|---|---|
| Ratio of stiffness | 1.5 | Static friction coefficient | 0.5 |
| Shear modulus(MPa) | 469 | Coefficient of rolling friction | 0.001 |

**Tab.2 BPM parameters**

| Normal stiffness (GN m$^{-3}$) | 1670 | Tangential critical pressure(MPa) | 24 |
|---|---|---|---|
| Tangential stiffness (GN m$^{-3}$) | 667 | Bonding radius(mm) | 0.8 |
| Normal critical pressure(MPa) | 36 | | |

### 3.2 Slop failure

When the metal plate focus on position B ($xB = -62.5$mm), the velocity vector of soil particles changes with the loading time. At the beginning of loading, the soil on both sides of the metal plate starts to move to the left and right sides due to the extrusion effect. The soil on the left side tends to uplift due to the fixed constraints on the left side, while the soil on the right side moves downward and to the right under the action of deadweight and lower pressure plate due to its proximity to the free end. With the continuous downward action of the metal plate, cracks and shear slip failure appear along the slope direction. When the loading is stopped, most of the soil particles gradually return to the stable state, and only a small amount of soil particles slide down from the slope accumulation under the action of gravity. Based on the change trend of the velocity vector of soil particles and the position of the slip crack surface, the validity of the discrete element model is verified.

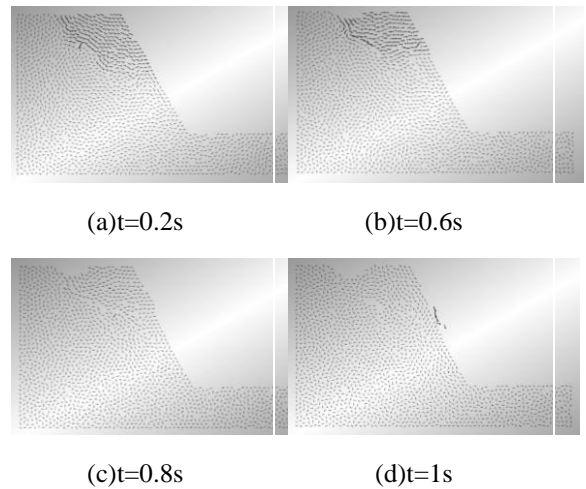

(a)t=0.2s  (b)t=0.6s

(c)t=0.8s  (d)t=1s

**Fig.2 slope particle velocity vector**

### 3.3 Effect of load position

In order to explore the effect of the load position on the slope failure, the discrete element simulation of the slope failure process at different positions of A, B, C, and D is carried out. The results are shown in Figure 3. It changes with the changing position of the metal plate. When loading position A, the angle of slope slipping and cracking surface is 63°, which is slightly larger than the angle of soil slope accumulation; when loading position B, the angle of slope slipping and cracking surface is 40°; When loading position at C, the angle of the slope slipping and cracking surface is 31°; when loading position at D, the slope soil is only squeezed directly below the loading plate, and the soil on both sides of the loading plate remains in a stable state, the tendency of relative motion is small, and there is no slippage and cracking surface.

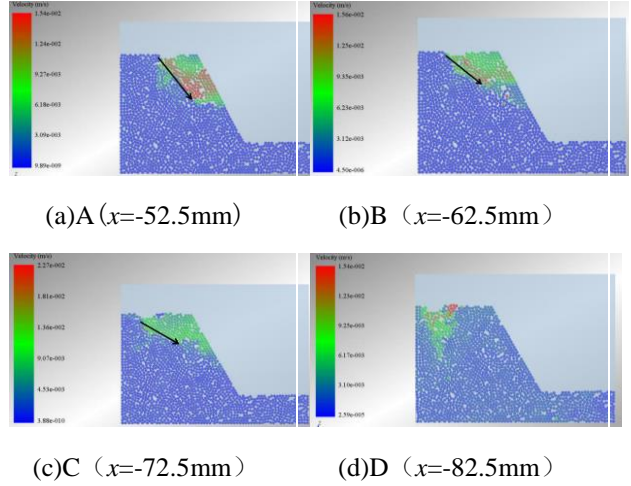

(a)A (*x*=-52.5mm)   (b)B (*x*=-62.5mm)

(c)C (*x*=-72.5mm)   (d)D (*x*=-82.5mm)

**Fig.3 Slope particle velocity contour at different action locations**

In the process of slope failure, the position of the slope top is often the first to deform. Combining the simulation effects of the previous four groups of slope failure, the average speed of the slope top measuring point *E* in the *y* direction is monitored in real time. Figure 4 is the *y*-direction average velocity-time curve of the slope top measuring point *E* at different acting positions. Compare the average velocity-time curve of the slope top measuring point *E* at the three different acting positions of A, B and C. As the position of the metal plate is farther from the measuring point *E* on the top of the slope, the peak of the upward movement velocity of the measuring point *E* appears later. This is because the farther the loading point is, the longer the transfer time of the compression effect of the metal lower plate on the soil; and as the position of the metal plate is farther away from the slope top measuring point *E*, the lower the peak of the downward movement speed of the measuring point *E* appears. This is because the farther the metal plate loading point is, the slope top ,the slippage and cracking surface of the soil appears later. From the average speed-time curve of the slope top measuring point E when the acting position is D, it can be seen that when the metal plate acting position is far enough from the slope top measuring point, the compression effect of the metal plate on the soil can be transmitted to the top of the slope, but there will be no slippage and cracking surface.

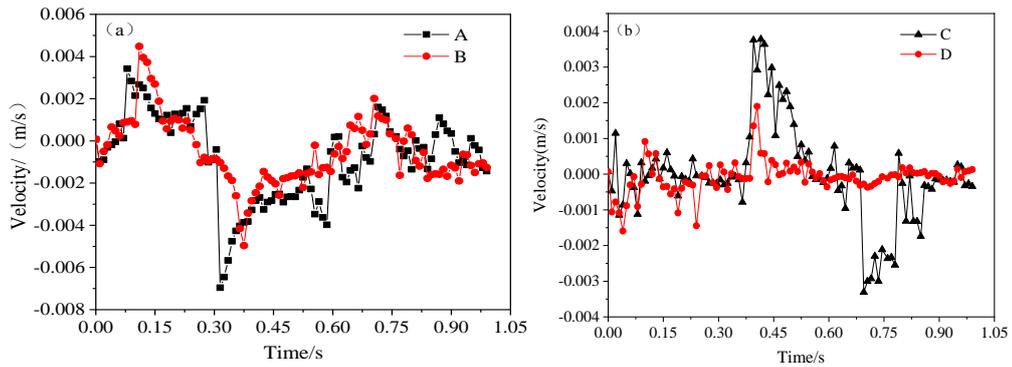

**Fig. 4 The average velocity-time curve of the *y*-direction of the Monitoring point E of the slope at different action locations**

## 4 Slope failure analysis based on persistent homology

This thesis uses the persistent homology theory to analyze the topological characteristics of the dynamic change of the structure network of the soil particle filling structure, quantify the evolution of the mesostructure of the slope soil particles under the external load, so as to better characterize the slope from static accumulation, soil cracking and then to the mechanical stage of slipping failure, the process of slope failure. Use JavaPlex software to calculate the topological characteristics of the force chain network of slope soil particles. Figure 5 shows the force chain network under the natural accumulation of slope soil particles. Since the slope top area is most affected when the slope is unstable, the force chain network in this area (frame selection) is selected for analysis, as shown in Fig.5.

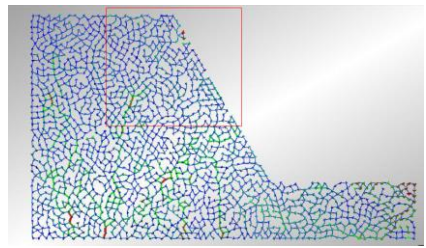

Fig.5 Force chain network under natural accumulation of slope soil particles

Observing the bar code diagram of the network at 0.2 seconds, in Fig.6, it is found that the 0-dimensional Betty number is divided into three levels as a whole. Most of the points are in the bottom layer, and its $\varepsilon$ is less than 1, which means that under the action of external load, the innermost particles are obviously tighter due to the compression of the left and right particles. The $\varepsilon$ of the middle layer point is slightly greater than 1, indicating that the particles are separated, the gap becomes larger, and the force chain disappears. Let's look at the 1-dimensional Betty number again. There are many 1-dimensional Betty numbers, that is, the force chain is tight, forming many 1-dimensional rings. This also shows that under the action of external load, the gaps between the particles are larger and larger holes are formed.

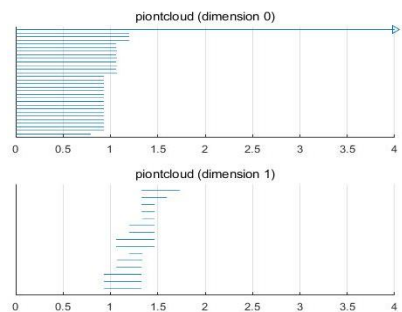

Fig.6 0.2 Second Barcode Chart

The 0-Dimension Betty number at 0.6s, in Fig.7, is more obvious and more layered than that at 0.2 s, and there are 4 layers. A few of them are less than 1, and most of them are greater than 1. This is obviously different from that of 0.2 s which also shows that with the continuous action of external load, more and more particles on the slope become loose and deviate from their original position, and the distance between particles is larger. 1-D Betty number is more sparse than 0.2 s, but there is a longer one. The results show that at 0.6 s, a slip band is formed due to the loosening of particles, and the force chain is damaged, and the remaining force chain is more sparse and distributed.

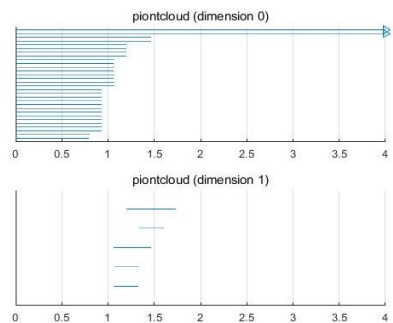

Fig.7 0.6 Second Barcode Chart

It can be found that the betty number of points with $\varepsilon$ greater than 1 in 0-dimensional increases rapidly and is more uneven, from the bar code diagram at 1s, in Fig.8, which indicates that greater displacement occurs at these points, that is, some soil particles at the outermost layer of the slope have completely separated.

The difference of 1-D Betty number between 1s and 0.6s is not obvious, it's still sparse, which indicates that the part of soil particles after the formation of slip band has a smaller impact on the force chain network due to the small slip between the soil particles and the external load on the upper surface, and the contact becomes weaker. Therefore, compared with 0.6s, the force chain network of this part has not changed much.

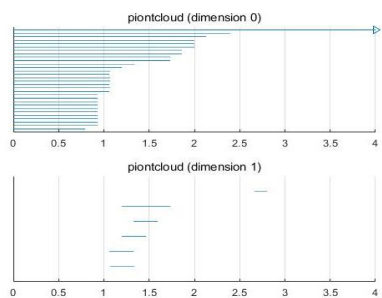

Fig.8 1 Second Barcode Chart.

To sum up, the 0-dimensional Betty number change trend of these three failure stages is from the beginning of regular and orderly with only one connector, to gradually becoming longer, becoming different in length, which indicates that the spacing of soil particles becomes larger, the force chain network becomes more dispersed, and each stage is consistent with the actual slope failure process. In addition, the bar code diagram of two adjacent stages has certain similarity, which shows that the failure process of the two adjacent stages is continuous, not jumping, which also conforms to the damage law of objects. The change trend of 1-D Betty number is that the initial line is shorter and denser, and the formation of 1-D holes is more and shorter. With the continuous action of load, the line becomes longer and more sparse. It shows that before the formation of slip surface, soil particles are closely connected under the action of external load. After the formation of slip surface, part of the external force is removed, so the holes formed by the force chain network become larger. At the same time, due to the fracture of part of the force chain caused by the slip surface, the one-dimensional Betty number becomes sparse.

## 5. Conclusion

From the above simulation and analysis, it can be seen that the results of persistent homology simulation are in good agreement with the real failure process of the slope. It can well explain the change process and topological characteristics of slope, so this method can be used in the identification of rock failure process in the future. The main conclusions are as follows.

There is a strong correlation between rock failure and its topological characteristics. If the 0-dimensional Betty number begins to lengthen significantly and becomes different in length, it indicates that the cracks in the rock mass structure are developing, the force chain network is deforming, and the rock mass is about to be destroyed.

If the ε of 1-D Betty number exceeds 1.5 and suddenly becomes longer, it indicates that the rock mass has formed a slip surface, and the sliding failure will occur.

With the help of the corresponding relationship between the rock mass failure and its topological characteristics bar code diagram established above, the development degree of internal cracks in rock mass can be obtained through X-ray and other means in the actual engineering rock slope, and then compared with the bar code diagram, then its failure trend can be predicted by the rock mass with a large degree of similarity.

## Acknowledgement

This work was supported by the Project Funded by Jiangxi Provincial Department of Science and Technology (No. 20192BBEL50028).

## References

[1] Li Ming-xia, Dong Lian-jie.Analysis on influential factors and deformation characteristics of toppling slope[J].Chinese Journal of Computational Mechanics, 2015(6):831-837. In Chinese
[2] Fakhimi A , Carvalho F , Ishida T , et al. Simulation of failure around a circular opening in rock[J]. International

Journal of Rock Mechanics & Mining Sciences, 2002, 39(4):507-515.

[3] Potyondy D O , Cundall P A . A bonded-particle model for rock[J]. International Journal of Rock Mechanics & Mining Sciences, 2004, 41(8):1329-1364.

[4] Backstrom A , Antikainen J , Backers T , et al. Numerical modelling of uniaxial compressive failure of granite with and without saline porewater[J]. International Journal of Rock Mechanics & Mining Sciences, 2008, 45(7):1126-1142.

[5] Hsieh Y M , Li H H , Huang T H , et al. Interpretations on how the macroscopic mechanical behavior of sandstone affected by microscopic properties—Revealed by bonded-particle model[J]. Engineering Geology, 2008, 99(1):1-10.

[6] Yang Bing. Particle dynamics simulation of slope dynamic failure process and collapse range [D]. Tsinghua University, 2011.In Chinese

[7] Zhang Xiaoxue. Slope stability analysis based on particle flow simulation [D]. 2015. In Chinese

[8] Yang Ling. Discrete element numerical simulation of loess collapse [D]. 2015. In Chinese

[9] Feng Chun, LI Shi-hai, SUN Hou-guang, et al. Particle contact meshless method and its application to simulation of slope disaster area[J]. Rock and Soil Mechanics, 2016, 37(12): 3608-3617. In Chinese

[10] Zou Yu. Multi-scale analysis of slope stability [D]. 2017. In Chinese

[11] Zhang Jing-liang,Ju Xian-men.Application of persistent homology in image classification and recognition[J].Journal of applied mathematics and computational mathematics,2017,(4)：494-508. In Chinese

[12] Quist J , Evertsson C M . Cone crusher modelling and simulation using DEM[J]. Minerals Engineering, 2015, 85:92-105.